\documentclass[12pt,reqno, twoside]{article}
\usepackage{amsfonts, amsmath, amssymb, amscd, amsthm}

\usepackage{multicol}
\usepackage{comment}
\usepackage{color} %Lin

 \theoremstyle{definition}
 
 \theoremstyle{remark}

 \numberwithin{equation}{section}

\numberwithin{equation}{section}

\newcounter{rom}
\renewcommand{\therom}{(\roman{rom})}
{\end{list}}

\title{The generalized Kupershmidt deformation for constructing new discrete integrable systems}

\author{Yehui Huang $^{a}$ \footnote{Corresponding author: huangyh@mails.tsinghua.edu.cn}, Runliang Lin $^{b}$ \footnote{rlin@math.tsinghua.edu.cn}, Yuqin Yao $^{c}$ \footnote{yyqinw@126.com} and Yunbo Zeng $^{b}$ \footnote{yzeng@math.tsinghua.edu.cn}}

\date{}

\begin{document}

\maketitle

$^{a}${\small\textit{School of Mathematics and Physics, North China Electric Power University, Beijing, 102206, China}}

$^{b}${\small\textit{Department of Mathematical Sciences, Tsinghua University, Beijing, 100084, China}}

$^{c}${\small\textit{Department of
  Applied Mathematics, China Agricultural University, Beijing, 100083, China}}

%%% ----------------------------------------------------------------------

\begin{abstract}
KdV6 equation can be described as the Kupershmidt deformation of the KdV equation (see {\it 2008, Phys. Lett. A 372: 263}).
In this paper, starting
from the bi-Hamiltonian structure of the discrete integrable system, we
propose a generalized Kupershmidt deformation to construct new discrete integrable
 systems.
Toda hierarchy, Kac-van Moerbeke
hierarchy and Ablowitz-Ladik hierarchy are considered.
The Lax representations for
these new deformed systems are presented. The generalized Kupershmidt deformation for the discrete integrable systems provides a new way to construct new discrete integrable systems.
\end{abstract}



%%% ----------------------------------------------------------------------
\maketitle
%%% ----------------------------------------------------------------------
%\tableofcontents
\section{Introduction}
In recent years,
 the deformations of integrable systems attracted lots of attention
 \cite{Kupershmidt}-\cite{Yao3}.
 It is known that one can construct a new integrable system from a bi-Hamiltonian system \cite{Fuchssteiner,Olver}. KdV6 equation is derived by A. Karasu-Kalkani,  {\it et al.,} \cite{Kalkanli} by means of the Painlev\'e analysis. KdV6 equation is written as
\begin{subequations}\label{KdV6}
\begin{align}
  &u_t=u_{xxx}+6uu_x-\omega_x,\\
  &\omega_{xxx}+4u\omega_x+2u_x\omega=0,
\end{align}
\end{subequations}
which could also be seen as the nonholonomic deformation of KdV equation \cite{Kalkanli}.
Many authors studied the integrability properties of KdV6,
such as zero-curvature representation, bi-Hamiltonian structure, conserved quantities, multisolitons and so on \cite{Kalkanli}-\cite{Yao}.
Kupershimidt found that \eqref{KdV6} can be converted into
\begin{subequations}\label{KdV6-2}
  \begin{align}
    &u_t=J(\frac{\delta H_3}{\delta u})-J(\omega),\\
    &K(\omega)=0,
  \end{align}
\end{subequations}
where
$$J=\partial=\partial_x,\quad K=\partial^3+2(u\partial+\partial u), $$
are two standard Hamiltonian operators of the KdV hierarchy and $H_3=u^3-\frac{u_x^2}{2}$ is the Hamiltonian function.
In general, for a bi-Hamiltonian system
$$u_{t_m}=J(\frac{\delta H_{m+1}}{\delta u})=K(\frac{\delta H_m}{\delta u}),$$
there is a nonholonomic deformation of bi-Hamiltonian system \cite{Kupershmidt}
\begin{subequations}\label{KdV6-3}
  \begin{align}
    &u_{t_m}=J(\frac{\delta H_{m+1}}{\delta u})-J(\omega),\\
    &K(\omega)=0,
  \end{align}
\end{subequations}
which is named as Kupershmidt deformation of bi-Hamiltonian systems.

In \cite{Yao}, we proved that the Kupershmidt deformation of the KdV equation could be seen as the Rosochatius deformation of the KdV equation with self-consistent sources,
and %Lin we obtained
 its bi-Hamiltonian structure are obtained.
We also proposed a generalized Kupershmidt deformation (GKD) for some bi-Hamiltonian system with pure differential Hamiltonian operators in \cite{Yao2} and \cite{Yao3}. For an integrable system,
we know that its two Hamiltonian operators have the relation
$$K(\frac{\delta\lambda_j}{\delta u})=\gamma_j J(\frac{\delta\lambda_j}{\delta u}),$$
where $\frac{\delta\lambda_j}{\delta u}$ can be taken from the spectral problem
of the system and $\gamma_j=\lambda_j$ or $\gamma_j=\lambda_j^2$ depending on
which integrable system we are studying.
For the integrable system with self-consistent sources, we have eigenfunctions and adjoint eigenfunctions as components of the sources and they will satisfy the spectral problem and the adjoint spectral problem corresponding to $\lambda_j$. In the GKD case, we assume that the nonholonomic term satisfy $(K-\gamma_j J)(\frac{\delta\lambda_j}{\delta u})=0$ instead of the spectral problem
so that they will have more freedom as they satisfy higher order equations. For the continuous case,
we have found that the GKD
of the integrable system will have the form \cite{Yao2,Yao3}
\begin{subequations}\label{Kupershmidt}
  \begin{align}
    &u_{t_m}=J(\frac{\delta H_{m+1}}{\delta u}-\sum_{j=1}^N\frac{\delta\lambda_j}{\delta u}),\\
    &(K-\gamma_j J)(\frac{\delta\lambda_j}{\delta u})=0, j=1,2,\ldots,N,
  \end{align}
\end{subequations}
where $\gamma_j$ equals to $\lambda_j$ or $\lambda_j^2$
 depending on the integrable system.

The integrability of the generalized system is proved by constructing its Lax representation.
The present paper focuses on
the GKD for the discrete integrable systems. We take
Toda lattice hierarchy,
Kac-van Moerbeke hierarchy and Ablowitz-Ladik hierarchy as examples
to construct the generalized Kupershmidt deformation for these systems
and %Lin we obtain
 their Lax representation are obtained to show that the new systems are integrable generalization of the original discrete systems.

Zhou introduced a concept of mixed hierarchy of soliton equations combining the negative order equations to the positive order equations and showed that the Kupershmidt deformed systems are just special members in the mixed soliton hierarchies \cite{Zhou}.
The mixed hierarchy of soliton equations \cite{Zhou},
 which has one nonholonomic term, are some special nonholonomic deformed
equations. Our GKD admit $N$ nonholonomic terms.

The paper is organized as follows. In section 2, we construct the GKD for  %Lin
the Toda hierarchy. Section 3 treats the GKD for  %Lin
 the Kac-van Moerbeke hierarchy. In Section 4, we will obtain the GKD for  %Lin
  the Ablowitz-Ladik hierarchy. Some conclusions will be given in Section 5.

\section{The generalized Kupershmidt deformation of Toda hierarchy}
\setcounter{equation}{0}
Let us recall some fundamental concepts
of the discrete integrable system. Assume $f=f(n,t)$ for $n\in\mathbb{Z}$ and $t\in\mathbb{R}$. The shift operator $E$ and difference operator $D$ are defined as
\begin{align}
  &(Ef)(n)=f(n+1),~(E^{-1}f)(n)=f(n-1),\notag\\
  &(Df)(n)=(E-1)f(n),~n\in\mathbb{Z},
\end{align}
and $f^{(j)}=E^jf,~j\in\mathbb{Z}$.

Consider the following discrete eigenvalue problem for the Toda hierarchy \cite{Newell}
\begin{equation}\label{toda-x}
  \Psi^{(-1)}=U(u,v,\lambda)\Psi,
\end{equation}
where
$$U(u,v,\lambda):=\left(\begin{array}{cc}0 & 1\\-v^{(1)} & \lambda-u\\\end{array}\right),~ \Psi=\left(\begin{array}{c} \psi^{(1)}\\ \psi \\\end{array}\right).$$
First,
 we consider the following stationary zero-curvature equation for generating function $\Gamma$,
\begin{equation}\label{generating}
    \Gamma^{(-1)}U-U\Gamma=0,
\end{equation}
where $$\Gamma=\sum_{i=0}^{+\infty}\Gamma_i\lambda^{-i}=\sum_{i=0}^{+\infty} \left(\begin{array}{cc} a_i & b_i \\ c_i & -a_i \\ \end{array}\right )\lambda^{-i},$$
and $a_i,~b_i,~c_i$ are functions of $n$ and $t$. In general, the recursion formula is
\begin{equation*}
    D a_{i+1}=u^{(1)}D a_i-c_i-b_i^{(1)}v^{(2)},~b_{i+1}=u^{(1)}b_i-a_i^{(1)}-a_i, ~c_i=-v^{(1)}b_i^{(-1)},
\end{equation*}
where the initial data is chosen as $a_0=\frac{1}{2},~b_0=c_0=0.$
Define the modification matrix
$$\Delta_m:=diag(b_{m+1}+\delta,\delta),$$
where $\delta$ is an arbitrary constant. Let
\begin{equation}\label{toda-t}
\Psi_{t_m}=V_m\Psi.
\end{equation}
where
$$V_m:=(\lambda^m\Gamma)_++\Delta_m=\sum_{i=0}^m\Gamma_i \lambda^{m-i}+\Delta_m.$$
Then the compatibility condition of \eqref{toda-x} and \eqref{toda-t}
gives rise to the zero-curvature representation of the Toda lattice hierarchy
\begin{equation}\label{toda-zero}
    U_{t_m}=UV_m-V_m^{(-1)}U.
\end{equation}
For $m=1$, we have the famous Toda equation
\begin{subequations}
\begin{align}\label{toda equ}
    &v_t=v(u^{(-1)}-u),\\
    &u_t=v-v^{(1)}
\end{align}
\end{subequations}
The bi-Hamiltonian structure of the toda hierarchy reads
(see, e.g., \cite{Lin-2002})
$$\left(\begin{array}{c} v \\ u \\\end{array}\right)_{t_m}=J^{(T)}\left(\begin{array}{c} a_{m+1}^{(-1)}/v \\ -b_{m+1}^{(-1)} \\\end{array}\right)=K^{(T)}\left(\begin{array}{c} a_{m}^{(-1)}/v \\ -b_{m}^{(-1)} \\\end{array}\right),$$
where
\begin{subequations}
\begin{align}\label{Toda ham}
    &J^{(T)}=\left(\begin{array}{cc}0 & v(E^{-1}-1)\\(1-E)v & 0\\\end{array}\right),\\
    &K^{(T)}=\left(\begin{array}{cc} -vEv+vE^{-1}v & -vu+vE^{-1}u \\ vu-uEv & vE^{-1}-Ev \\ \end{array} \right),
\end{align}
\end{subequations}
are two standard Hamiltonian operators.

Now we derive $\frac{\delta \lambda_j}{\delta v}$ and $\frac{\delta \lambda_j}{\delta u}$ from \eqref{toda-x} and its adjoint problem with $\lambda=\lambda_j$. Define $L=v^{(1)}E+u+E^{-1}$ and its adjoint operator is $L^*=vE^{-1}+u+E$. Then we have $L\psi_j=\lambda_j\psi_j$, $L^*\phi_j=\lambda_j\phi_j$. We assume that $L$, $\lambda_j$, $\psi_j$ and $\phi_j$ have a perturbation $\varepsilon$. So we can assume that they are functions of $\varepsilon$. Furthermore we can properly choose $\phi_j$ to make $<\psi_j,\phi_j>=1$, where the inner product is defined as $<f,g>=\sum_n f(n)g(n)$. With all these assumptions we have
\begin{align}
  &\frac{d}{d\varepsilon}(\lambda_j(\varepsilon))=\frac{d}{d\varepsilon}(<\lambda_j\psi_j,\phi_j>)=\frac{d}{d\varepsilon}(<L\psi_j,\phi_j>)\notag\\
  &=<\frac{dL}{d\varepsilon}\psi_j,\phi_j>+<L\frac{d\psi_j}{d\varepsilon},\phi_j> +<L\psi_j,\frac{d\phi_j}{d\varepsilon}>\notag\\
  &=<\frac{dv^{(1)}}{d\varepsilon}\psi_j^{(1)}+\frac{du}{d\varepsilon}\psi_j,\phi_j>+<\frac{d\psi_j}{d\varepsilon},L^*\phi_j> +<L\psi_j,\frac{d\phi_j}{d\varepsilon}>\notag\\
  &=<\frac{dv^{(1)}}{d\varepsilon}\psi_j^{(1)}+\frac{du}{d\varepsilon}\psi_j,\phi_j>+\lambda_j\frac{d}{d\varepsilon}(<\psi_j,\phi_j>)\notag\\
  &=<\frac{dv^{(1)}}{d\varepsilon}\psi_j^{(1)}+\frac{du}{d\varepsilon}\psi_j,\phi_j>.\notag
\end{align}

So we have
$$\left(\begin{array}{c} \frac{\delta \lambda_j}{\delta v} \\ \frac{\delta \lambda_j}{\delta u} \\\end{array}\right)=\left(\begin{array}{c} \psi_j\phi_j^{(-1)} \\ \psi_j\phi_j \\\end{array}\right),$$

Take $\gamma_j=\lambda_j$, \eqref{Kupershmidt} gives rise to the following new generalized Toda lattice hierarchy
\begin{subequations}
  \begin{align}
  &\left(\begin{array}{c} v \\ u \\\end{array}\right)_{t_m}=J^{(T)}(\left(\begin{array}{c} a_{m+1}^{(-1)}/v \\ -b_{m+1}^{(-1)} \\\end{array}\right)-\sum_{j=1}^N\left(\begin{array}{c} \bar\psi_j\bar\phi_j^{(-1)} \\ \bar\psi_j\bar\phi_j \\\end{array}\right)), \label{Todanew-a} \\
  &(K^{(T)}-\lambda_j J^{(T)})\left(\begin{array}{c} \bar\psi_j\bar\phi_j^{(-1)} \\ \bar\psi_j\bar\phi_j \\\end{array}\right)=0, \qquad j=1,2,\ldots,N. \label{Todanew-b}
  \end{align}
\end{subequations}
It is noted that in our new deformed system, $\bar\psi_j$ and $\bar\phi_j$ are no longer the eigenfunctions and adjoint eigenfunctions of the spectral problem, but are solutions of \eqref{Todanew-b}. After simplifying \eqref{Todanew-b}, we can find that
\begin{subequations}
  \begin{align}
    &-v^{(1)}\bar\psi_j^{(1)}\bar\phi_j+v^{(-1)}\bar\psi_j^{(-1)}\bar\phi_j^{(-2)}-u\bar\psi_j\bar\phi_j \notag\\ &+u^{(-1)}\bar\psi_j^{(-1)}\bar\phi_j^{(-1)}-\lambda_j\bar\psi_j^{(-1)}\bar\phi_j^{(-1)} +\lambda_j\bar\psi_j\bar\phi_j=0,\\
    &vu\bar\psi_j\bar\phi_j^{(-1)}-uv^{(1)}\bar\psi_j^{(1)}\bar\phi_j+v\bar\psi_j^{(-1)}\bar\phi_j^{(-1)} \notag\\ &-v^{(1)}\bar\psi_j^{(1)}\bar\phi_j^{(1)}+\lambda_jv^{(1)}\bar\psi_j^{(1)}\bar\phi_j -\lambda_jv\bar\psi_j\bar\phi_j^{(-1)}=0.
  \end{align}
\end{subequations}
Denote $f_j=u\bar\psi_j+v^{(1)}\bar\psi_j^{(1)}+\bar\psi_j^{(-1)}-\lambda_j\bar\psi_j$
and $g_j=u\bar\phi_j+v\bar\phi_j^{(-1)}+\bar\phi_j^{(1)}-\lambda_j\bar\phi_j$, the above equations will lead to
\begin{subequations}
  \begin{align}
    &\bar\phi_jf_j-\bar\psi_j^{(-1)}g_j^{(-1)}=0,\\
    &v\bar\phi_j^{(-1)}f_j-v^{(1)}\bar\psi_j^{(1)}g_j=0,
  \end{align}
\end{subequations}
which gives that\begin{subequations}
  \begin{align}
    &f_j=\frac{\mu_j}{v\bar\psi_j\bar\phi_j\bar\phi_j^{(-1)}},\\
    &g_j=\frac{\mu_j}{v^{(1)}\bar\psi_j\bar\psi_j^{(1)}\bar\phi_j}.
  \end{align}
\end{subequations}
where $\mu_j$ are some arbitrary constants.
Therefore, we make the simplification of \eqref{Todanew-b} and find that our new constraints are the original spectral problem with some addition terms with some arbitrary constants $\mu_j$.
the GKD of Toda lattice hierarchy is
\begin{subequations}
  \begin{align}
  &\left(\begin{array}{c} v \\ u \\\end{array}\right)_{t_m}=J^{(T)}(\left(\begin{array}{c} a_{m+1}^{(-1)}/v \\ -b_{m+1}^{(-1)} \\\end{array}\right)-\sum_{j=1}^N\left(\begin{array}{c} \bar\psi_j\bar\phi_j^{(-1)} \\ \bar\psi_j\bar\phi_j \\\end{array}\right)), \\
  &u\bar\psi_j+v^{(1)}\bar\psi_j^{(1)}+\bar\psi_j^{(-1)}=\lambda_j\bar\psi_j+\frac{\mu_j}{v\bar\psi_j\bar\phi_j\bar\phi_j^{(-1)}},\\
  &u\bar\phi_j+v\bar\phi_j^{(-1)}+\bar\phi_j^{(1)}=\lambda_j\bar\phi_j+\frac{\mu_j}{v^{(1)}\bar\psi_j\bar\psi_j^{(1)}\bar\phi_j}.
  \end{align}
\end{subequations}
We can find that in the GKD of the Toda hierarchy, the constraints is the spectral problem and its adjoint problem with some nonholonomic terms with arbitrary constants $\mu_j$.

When $m=1$, the GKD of Toda equation reads
\begin{subequations}\label{GKDtodae}
  \begin{align}
    &v_t=v(u^{(-1)}+\sum_{j=1}^N\bar\psi_j^{(-1)}\bar\phi_j^{(-1)})-v(u+\sum_{j=1}^N\bar\psi_j\bar\phi_j),\\
    &u_t=v(1+\sum_{j=1}^N\bar\psi_j\bar\phi_j^{(-1)})-v^{(1)}(1+\sum_{j=1}^N\bar\psi_j^{(1)}\bar\phi_j),\\
    &u\bar\psi_j+v^{(1)}\bar\psi_j^{(1)}+\bar\psi_j^{(-1)}=\lambda_j\bar\psi_j+\frac{\mu_j}{v\bar\psi_j\bar\phi_j\bar\phi_j^{(-1)}},\label{todacons-a}\\
    &u\bar\phi_j+v\bar\phi_j^{(-1)}+\bar\phi_j^{(1)}=\lambda_j\bar\phi_j+\frac{\mu_j}{v^{(1)}\bar\psi_j\bar\psi_j^{(1)}\bar\phi_j}.\label{todacons-b}
  \end{align}
\end{subequations}
 and
its Lax representation is
\begin{subequations}\label{GKDtodae-Lax}
  \begin{align}
    &(v^{(1)}E+u+E^{-1})\psi=\lambda\psi,\\
    &-\psi_t=v^{(1)}\psi^{(1)}+\sum_{j=1}^{N}\frac{v^{(1)}\bar\phi_j}{\lambda-\lambda_j}(\psi\bar\psi_j^{(1)}-\psi^{(1)}\bar\psi_j).
  \end{align}
\end{subequations}
 The compatibility of the system (\ref{GKDtodae-Lax})
under the condition \eqref{todacons-a} and \eqref{todacons-b}
will give (\ref{GKDtodae}a) and (\ref{GKDtodae}b).
It is an interesting question to find a Lax pair with $\mu_j$ inside
to give the system (\ref{GKDtodae}).
This will be studied in future.
%Lin This shows that our new generalized system is Lax integrable.
%Lin We would find that the Lax pair doesn't have terms with $\mu_j$ because when we consider the compatibility of the Lax pair under the constraints \eqref{todacons-a} and \eqref{todacons-b}, we would find that the terms with $\mu_j$ eliminates after the calculation.
When we take $\lambda_j=0$, \eqref{GKDtodae} reduces to the Kupershmidt deformation of the Toda equation. When we take $\mu_j=0$, \eqref{GKDtodae}  reduces to the Toda equation
with self-consistent sources \cite{Liu}.

\section{The generalized Kupershmidt deformation of Kac-van Moerbeke hierarchy}
Consider the eigenvalue problem for the Kac-van Moerbeke hierarchy \cite{Zeng}
\begin{equation}
  (E+vE^{-1})\psi=\lambda\psi.
\end{equation}
Its adjoint equation reads
\begin{equation}
  (E^{-1}+Ev)\phi=\lambda\phi.
\end{equation}
We can also write the eigenvalue problem in matrix form
\begin{equation}\label{Kvm space}
  E\left(\begin{array}{c} \psi^{(-1)} \\ \psi \end{array} \right)=U \left(\begin{array}{c} \psi^{(-1)} \\ \psi \end{array} \right),~U=\left( \begin{array}{cc} 0 & 1 \\ -v & \lambda \end{array} \right).
\end{equation}
To derive the hierarchy of evolution equations associated with the eigenvalue problem, we frst solve the stationary discrete zero-curvature equation
\begin{equation}\label{Kvm zero}
    \Gamma^{(1)}U-U\Gamma=0.
\end{equation}
Let
\begin{equation}\label{Kvm Gamma}
    \Gamma=\left(\begin{array}{cc} a & b \\ c & -a \\ \end{array} \right)=\sum_{i=0}^{+\infty}\Gamma_i\lambda^{-i} =\sum_{i=0}^{+\infty}\left(\begin{array}{cc} a_i & b_i \\ c_i & -a_i \\ \end{array} \right)\lambda^{-i}
\end{equation}
and take the initial data as
\begin{equation}\label{Kvm ini}
    a_0=\frac{1}{2},~~b_0=0,~~b_1=-1.
\end{equation}
We can find the recursion formula
\begin{align*}
    &b_{2i}=c_{2i}=a_{2i+1}=0,~~i=0,1,\ldots\\
    &c_{2i+1}=-vb_{2i+1}^{(1)},\\
    &b_{2i+1}=-(a_{2i}^{(-1)}+a_{2i}),\\
    &c_{2i+1}^{(1)}+vb_{2i+1}=a_{2i+2}^{(1)}-a_{2i+2}.
\end{align*}
Set the auxiliary linear problem as
\begin{equation}\label{Kvm time}
    \psi_{t_m}=V_{2m}\psi
\end{equation}
with
\begin{align}\label{Kvm V}
    &V_{2m}=(\Gamma\lambda^{2m})_++\Delta_{2m}\notag\\
    &=\left( \begin{array}{cc} \sum_{i=0}^{m}a_{2i}\lambda^{2m-2i} & \sum_{i=0}^{m-1}b_{2i+1}\lambda^{2m-2i-1} \\ \sum_{i=0}^{m-1}c_{2i+1}\lambda^{2m-2i-1} & -\sum_{i=0}^{m}a_{2i}\lambda^{2m-2i} \\\end{array} \right) + \left( \begin{array}{cc} b_{2m+1} & 0 \\ 0 & 0 \\\end{array} \right).
\end{align}
Then the compatibility condition of \eqref{Kvm space} and \eqref{Kvm time} gives rise to the zero-curvature representation of the Kac-van Moerbeke hierarchy
\begin{equation}\label{kvm zero}
    U_{t_m}=V_{2m}^{(1)}U-UV_{2m}
\end{equation}
When $m=1$, we have the Kac-van Moerbeke equation
\begin{equation}\label{Kvm}
    v_t=v(v^{(-1)}-v^{(1)}).
\end{equation}
The Hamiltonian operators of Kac-van Moerbeke hierarchy are defined as
\begin{subequations}
  \begin{align}
    &J^{(KvM)}=v(E^{-1}-E)v,\\
    &K^{(KvM)}=v(vE^{-1}+v^{(-1)}E^{-1}+v^{(-1)}E^{-2}-vE-v^{(1)}E^{2}-v^{(1)}E)v.
  \end{align}
\end{subequations}
The Kac-van Moerbeke hierarchy has the bi-Hamiltonian form
\begin{equation}
v_{t_m}=J^{(KvM)}\frac{\delta H_{2m}}{\delta v}=K^{(KvM)}\frac{ H_{2m-2}}{\delta v},
\end{equation}
where $H_{2m}=-\frac{b_{2m+1}}{2m}$ is the Hamiltonian function.
It is not difficult to find that
\begin{equation}\label{Kvm deltav}
    \frac{\delta\lambda_j}{\delta v}=\psi_j^{(-1)}\phi_j.
\end{equation}

The new generalized Kac-van Moerbeke hierarchy is constructed as follows
\begin{subequations}
  \begin{align}
    &v_{t_m}=J^{(KvM)}\frac{\delta H_{2m}}{\delta v}+J^{(KvM)}\sum_{j=1}^N(\bar\psi_j^{(-1)}\bar\phi_j),\label{Kvm-a}\\
    &(K^{(KvM)}-\lambda_j^2J^{(KvM)})(\bar\psi_j^{(-1)}\bar\phi_j)=0.\label{Kvm-b}
  \end{align}
\end{subequations}
We assume that
\begin{subequations}
\begin{align}
    &(E+vE^{-1})\bar\psi_j=\lambda_j\bar\psi_j+f_j,\label{Kvm f}\\
    &(E^{-1}+Ev)\bar\phi_j=\lambda_j\bar\phi_j+g_j.\label{Kvm g}
\end{align}
\end{subequations}
By simplifying \eqref{Kvm-b} we have
\begin{align}\label{Kvm-b2}
  &(vv^{(-1)}\bar\psi_j^{(-2)}-v^{(1)}\bar\psi_j)g_j+(v^{(-1)}\bar\phi_j^{(-1)}-vv^{(1)}\bar\phi_j^{(1)})f_j^{(-1)}\notag\\
  &+v^{(-1)}(\lambda_j\bar\psi_j^{(-2)}-\bar\psi_j^{(-1)})g_j^{(-1)}+v^{(-1)}(\lambda_j\bar\phi_j^{(-1)}-v\bar\phi_j)f_j^{(-2)}\notag\\
  &-v^{(1)}(\lambda_j\bar\psi_j-v\bar\psi_j^{(-1)})g_j^{(1)}-v^{(1)}(\lambda_j\bar\phi_j^{(1)}-\bar\phi_j)f_j\notag\\
  &+v^{(-1)}g_j^{(-1)}f_j^{(-2)}-v^{(1)}g_j^{(1)}f_j+(v^{(1)}-v^{(-1)})(\bar\psi_j\bar\phi_j^{(-1)}-v\bar\psi_j^{(-1)}\bar\phi_j)=0.
\end{align}
 To calculate \eqref{Kvm f}$\times\bar\phi_j-$\eqref{Kvm g}$\times\bar\psi_j$, we will find that
\begin{equation*}
    \bar\phi_jf_j-\bar\psi_jg_j=(E-1)(\bar\psi_j\bar\phi_j^{(-1)}-v\bar\psi_j^{(-1)}\bar\phi_j).
\end{equation*}
Therefore we make the assumption that
\begin{equation}\label{Kvm ass1}
    v\bar\phi_j^{(-1)}f_j^{(-1)}=v\bar\psi_jg_j,
\end{equation}
which will lead to
\begin{equation}\label{Kvm psig}
    \bar\psi_jg_j=\bar\psi_j\bar\phi_j^{(-1)}-v\bar\psi_j^{(-1)}\bar\phi_j.
\end{equation}
Substituting \eqref{Kvm psig} into \eqref{Kvm-b2} and with
the assumption
\begin{equation}\label{Kvm ass2}
    v^{(1)}\bar\phi_j^{(1)}f_j=v\bar\psi_j^{(-1)}g_j,
\end{equation}
we will find \eqref{Kvm-b2} holds.
With the assumption \eqref{Kvm ass1} and \eqref{Kvm ass2}, we can solve for $f_j$ and $g_j$ as
\begin{subequations}\label{Kvm fgsolve}
\begin{align}
&f_j=\frac{\mu_j}{v^{(1)}\bar\psi_j\bar\phi_j\bar\phi_j^{(1)}},\\
&g_j=\frac{\mu_j}{v\bar\psi_j^{(-1)}\bar\psi_j\bar\phi_j}.
\end{align}
\end{subequations}
where $\mu_j$ are some arbitrary constants. So the GKD of Kac-van Moerbeke hierarchy is
\begin{subequations}\label{Kvmhgkd}
  \begin{align}
    &v_{t_m}=J^{(KvM)}\frac{\delta H_{2m}}{\delta v}+J^{(KvM)}\sum_{j=1}^N(\bar\psi_j^{(-1)}\bar\phi_j),\\
    &\bar\psi_j^{(1)}+v\bar\psi_j^{(-1)}=\lambda_j\bar\psi_j+\frac{\mu_j}{v^{(1)}\bar\psi_j\bar\phi_j\bar\phi_j^{(1)}},\\
    &\bar\phi_j^{(-1)}+v^{(1)}\bar\phi_j^{(1)}=\lambda_j\bar\phi_j+\frac{\mu_j}{v\bar\psi_j^{(-1)}\bar\psi_j\bar\phi_j}.
  \end{align}
\end{subequations}
When $m=1$, the GKD of Kac-van Moerbeke equation is
\begin{subequations}\label{Kvmgkd}
  \begin{align}
    &v_t=v(v^{(-1)}-v^{(1)})+\sum_{j=1}^N(vv^{(-1)}\bar\psi_j^{(-2)}\bar\phi_j^{(-1)}-vv^{(1)}\bar\psi_j\bar\phi_j^{(1)}),\\
    &\bar\psi_j^{(1)}+v\bar\psi_j^{(-1)}=\lambda_j\bar\psi_j+\frac{\mu_j}{v^{(1)}\bar\psi_j\bar\phi_j\bar\phi_j^{(1)}},\\
    &\bar\phi_j^{(-1)}+v^{(1)}\bar\phi_j^{(1)}=\lambda_j\bar\phi_j+\frac{\mu_j}{v\bar\psi_j^{(-1)}\bar\psi_j\bar\phi_j}.
  \end{align}
\end{subequations}
which has the Lax pair
\begin{subequations}\label{Kvmgkd-Lax}
  \begin{align}
    &\psi^{(1)}+v\psi^{(-1)}=\lambda\psi,\\
    &\psi_t=\lambda v\psi^{(-1)}-(\frac{1}{2}\lambda^2+v)\psi\notag\\
    &~~~~~~+\sum_{j=1}^N\frac{\lambda_j^2}{\lambda^2-\lambda_j^2}v\bar\psi_j^{(-1)}\bar\phi_j\psi-\frac{\lambda\lambda_j}{\lambda^2-\lambda_j^2}v\bar\psi_j\bar\phi_j\psi^{(-1)}.
  \end{align}
\end{subequations}
The compatibility of the system (\ref{Kvmgkd-Lax})
under the condition (\ref{Kvmgkd}c) and (\ref{Kvmgkd}d)
will give (\ref{Kvmgkd}a) and (\ref{Kvmgkd}b).
It is also an interesting question to find a Lax pair with $\mu_j$ inside
to give the system (\ref{Kvmgkd}).
This will be studied in future.
%Lin This shows that our new generalized system is Lax integrable.
%Lin Similar with the Toda hierarchy, the Lax pair doesn't contain terms with $\mu_j$.
When we take $\lambda_j=0$,
the GKD of Kac-van Moerbeke equation
\eqref{Kvmgkd}
will reduce to the Kupershmidt deformation version. When
 taking $\mu_j=0$, \eqref{Kvmgkd}
 will lead to the Kac-van Moerbeke equation with self-consistent sources.

\section{The generalized Kupershmidt deformation of Ablowitz-Ladik hierarchy}
Consider the following Ablowitz-Ladik discrete isospectral problem\cite{AL}:
\begin{equation}\label{al-spec}
  E\psi=U\psi,~~U=\left(\begin{array}{cc} z & Q \\ R & 1/z \end{array}\right),~~\psi=(\psi_1,\psi_2)^T,
\end{equation}
where $Q=Q(n,t),~R=R(n,t)$, $z$ is the spectral parameter.
Its adjoint problem reads
\begin{equation}
  E^{(-1)}\phi=\phi U,~~\phi=(\phi_1,\phi_2),
\end{equation}
To derive the Ablowitz-Ladik hierarchy, we need first solve the discrete zero-curvature equation
\begin{equation}
  (E\Gamma)U-U\Gamma=0
\end{equation}
Let
\begin{equation}\label{AL-gamma}
  \Gamma=\left(\begin{array}{cc} A & B \\ C & -A \end{array} \right),
\end{equation}
we find that
\begin{subequations}\label{AL-rec}
  \begin{align}
    &A^{(1)}z+B^{(1)}R-Az-CQ=0,\\
    &A^{(1)}Q+B^{(1)}\frac{1}{z}-Bz+AQ=0,\\
    &C^{(1)}z-A^{(1)}R-AR-C\frac{1}{z}=0,\\
    &C^{(1)}Q-A^{(1)}\frac{1}{z}-BR+A\frac{1}{z}=0.
  \end{align}
\end{subequations}
We can expand the above relations in power series of $z$ and $\frac{1}{z}$, respectively\cite{ALzeng}.

\subsection{$\Gamma$ expanded in power series of $1/z$}
Assume
\begin{equation}
  \Gamma=\sum_{i=0}^{\infty}\left(\begin{array}{cc} A_{2i}z^{-2t} & B_{2i+1}z^{-2i-1} \\ C_{2i+1}z^{-2i-1} & -A_{2i}z^{-2i} \end{array} \right).
\end{equation}
Then, the recursion relation \eqref{AL-rec} leads to
\begin{subequations}
  \begin{align}
    &A_0^{(1)}-A_0=0~~B_1=Q(A_0^{(1)}+A_0)~~C_1^{(1)}=R(A_0^{(1)}+A_0),\\
    &A_{2i}^{(1)}-A_{2i}=QC_{2i-1}-RB_{2i-1}^{(1)}=QC_{2i+1}^{(1)}-RB_{2i+1},\\
    &B_{2i+1}=Q(A_{2i}^{(1)}+A_{2i})+B_{2i-1}^{(1)},\\
    &C_{2i+1}^{(1)}=R(A_{2i}^{(1)}+A_{2i})+C_{2i-1},~~i=1,2,\cdots
  \end{align}
\end{subequations}
where the initial value can be chosen as $A_0=\frac{1}{2}$.

The Ablowitz-Ladik hierarchy can be written in the Hamiltonian form\cite{ALzeng}
\begin{equation}
  \left(\begin{array}{c} Q \\ R \end{array}\right)_{t_m}=J^{(AL)}\frac{\delta H_m}{\delta u}=K^{(AL)}\frac{\delta H_{m-1}}{\delta u}
\end{equation}
where $H_m=-\frac{A_{2m}}{m}$ and the Hamiltonian operators are
\begin{subequations}\label{al-ham-1}
  \begin{align}
    &J^{(AL)}=\left(\begin{array}{cc} 0 & 1-RQ \\ RQ-1 & 0 \end{array}\right),\\
    &K^{(AL)}=\left(\begin{array}{cc} K_{11} & K_{12} \\ K_{21} & K_{22} \end{array}\right),\\
    &K_{11}=QD^{-1}Q^{(1)}(1-RQ)+(1-RQ)Q^{(1)}D^{-1}EQ,\notag\\
    &K_{12}=E(1-RQ)-QD^{-1}R^{(1)}E^{(2)}(1-RQ)-(1-RQ)Q^{(1)}D^{-1}ER,\notag\\
    &K_{21}=-E^{(-1)}(1-RQ)-RD^{(-1)}QE^{(-1)}(1-RQ)+(RQ-1)R^{(-1)}D^{-1}Q,\notag\\
    &K_{22}=RD^{-1}RE(1-RQ)+(1-RQ)R^{(-1)}D^{-1}R.\notag
  \end{align}
\end{subequations}
It is not difficult to find that $\frac{\delta z}{\delta Q}=\psi_2\phi_1,~\frac{\delta z}{\delta R}=\psi_1\phi_2$.
The GKD of Ablowitz-Ladik hierarchy is
\begin{subequations}\label{al-gkd}
  \begin{align}\label{al-gkd-a}
    &\left(\begin{array}{c} Q \\ R \end{array}\right)_{t_m}=J^{(AL)}\left(\begin{array}{c} \delta H_m/\delta Q-\sum_{j=1}^N\tilde\psi_{2,j}\tilde\phi_{1,j} \\ \delta H_m/\delta R-\sum_{i=1}^N\tilde\psi_{1,j}\tilde\phi_{2,j} \end{array}\right),\\\label{al-gkd-b}
    &(K^{(AL)}-z_j^2J^{(AL)})\left(\begin{array}{c} \tilde\psi_{2,j}\tilde\phi_{1,j} \\ \tilde\psi_{1,j}\tilde\phi_{2,j} \end{array}\right)=0.
  \end{align}
\end{subequations}
We assume that
\begin{subequations}\label{al-constraint-1}
  \begin{align}
    &E\tilde\psi_{1,j}=z_j\tilde\psi_{1,j}+Q\tilde\psi_{2,j}+f_{1,j},\\
    &E\tilde\psi_{2,j}=R\tilde\psi_{1,j}+\frac{1}{z_j}\tilde\psi_{2,j}+f_{2,j},\\
    &E^{(-1)}\tilde\phi_{1,j}=z_j\tilde\phi_{1,j}+R\tilde\phi_{2,j}+g_{1,j},\\
    &E^{(-1)}\tilde\phi_{2,j}=Q\tilde\phi_{1,j}+\frac{1}{z_j}\tilde\phi_{2,j}+g_{2,j}.
  \end{align}
\end{subequations}
Simplifying \eqref{al-gkd-b} we get
\begin{subequations}
  \begin{align}
    &QD^{-1}Q^{(1)}(1-RQ)\tilde\psi_{2,j}\tilde\phi_{1,j}+(1-RQ)Q^{(1)}D^{-1}EQ\tilde\psi_{2,j}\tilde\phi_{1,j}\notag\\ &+E(1-RQ)\tilde\psi_{1,j}\tilde\phi_{2,j}-QD^{-1}R^{(1)}E^{(2)}(1-RQ)\tilde\psi_{1,j}\tilde\phi_{2,j}\notag\\ &-(1-RQ)Q^{(1)}D^{-1}ER\tilde\psi_{1,j}\tilde\phi_{2,j}-z_j^2(1-RQ)\tilde\psi_{1,j}\tilde\phi_{2,j}=0,\\
    &E^{(-1)}(1-RQ)\tilde\psi_{2,j}\tilde\phi_{1,j}+RD^{-1}QE^{(-1)}(1-RQ)\tilde\psi_{2,j}\tilde\phi_{1,j}\notag\\ &+(1-RQ)R^{(-1)}D^{-1}Q\tilde\psi_{2,j}\tilde\phi_{1,j}-RD^{-1}RE(1-RQ)\tilde\psi_{1,j}\tilde\phi_{2,j}\notag\\ &-(1-RQ)R^{(-1)}D^{-1}R\tilde\psi_{1,j}\tilde\phi_{2,j}-z_j^2(1-RQ)\tilde\psi_{2,j}\tilde\phi_{1,j}=0.
  \end{align}
\end{subequations}
Substituting \eqref{al-constraint-1} into the above equation we find that
\begin{subequations}
  \begin{align}
    &z_j(1-RQ)(\tilde\phi_{2,j}f_{1,j}-\tilde\psi_{1,j}^{(1)}g_{2,j}^{(1)}) +(1-RQ)Q^{(1)}[\tilde\psi_{1,j}^{(1)}g_{1,j}^{(1)}+D^{-1}E(\tilde\psi_{1,j}g_{1,j}-\tilde\phi_{1,j}f_{1,j})]\notag\\ &+QD^{-1}[(1-RQ)\tilde\psi_{2,j}(Q^{(1)}g_{1,j}^{(1)}-zg_{2,j}^{(1)})-(1-RQ)^{(1)}\tilde\phi_{2,j}^{(1)}(Rf_{1,j}-z_jf_{2,j})]=0,\\
    &z_j(1-RQ)(\tilde\psi_{2,j}g_{1,j}-\tilde\phi_{1,j}^{(-1)}f_{2,j}^{(-1)}+(1-RQ)R^{(-1)}[\tilde\phi_{1,j}^{(-1)}f_{1,j}^{(-1)} +D^{-1}(\tilde\psi_{1,j}g_{1,j}-\tilde\phi_{1,j}f_{1,j})]\notag\\ &+RD^{-1}[(1-RQ)\tilde\psi_{2,j}(Q^{(1)}g_{1,j}^{(1)}-a_jg_{2,j}^{(1)})-(1-RQ)^{(1)}\tilde\phi_{2,j}^{(1)}(Rf_{1,j}-z_jf_{2,j})]=0.
  \end{align}
\end{subequations}
First we assume that
\begin{subequations}
  \begin{align}
    &f_{2,j}=\frac{R}{z_j}f_{1,j},\\
    &g_{2,j}=\frac{Q}{z_j}g_{1,j},
  \end{align}
\end{subequations}
then the above relations are simplified and
we can
find that
\begin{subequations}
  \begin{align}
    &f_{1,j}=\frac{\mu_j}{\tilde\phi_{1,j}-\frac{z_j\tilde\phi_{2,j}}{Q^{(-1)}}},\\
    &g_{1,j}=\frac{\mu_j}{\tilde\psi_{1,j}-\frac{z_j\tilde\psi_{2,j}}{R^{(-1)}}},
  \end{align}
\end{subequations}
where $\mu_j$ are some arbitrary constants.
So the GKD of the Ablowitz-Ladik hierarchy reads
\begin{subequations}\label{Al-system-1}
  \begin{align}
    &\left(\begin{array}{c} Q \\ R \end{array}\right)_{t_m}=J^{(AL)}\left(\begin{array}{c} \delta H_m/\delta Q-\sum_{j=1}^N\tilde\psi_{2,j}\tilde\phi_{1,j} \\ \delta H_m/\delta R-\sum_{i=1}^N\tilde\psi_{1,j}\tilde\phi_{2,j} \end{array}\right),\\
    &E\tilde\psi_{1,j}=z_j\tilde\psi_{1,j}+Q\tilde\psi_{2,j}+\frac{\mu_j}{\tilde\phi_{1,j}-\frac{z_j\tilde\phi_{2,j}}{Q^{(-1)}}},\\
    &E\tilde\psi_{2,j}=R\tilde\psi_{1,j}+\frac{1}{z_j}\tilde\psi_{2,j}+\frac{\mu_j R}{z_j(\tilde\phi_{1,j}-\frac{z_j\tilde\phi_{2,j}}{Q^{(-1)}})},\\
    &E^{(-1)}\tilde\phi_{1,j}=z_j\tilde\phi_{1,j}+R\tilde\phi_{2,j}+\frac{\mu_j}{\tilde\psi_{1,j}-\frac{z_j\tilde\psi_{2,j}}{R^{(-1)}}},\\
    &E^{(-1)}\tilde\phi_{2,j}=Q\tilde\phi_{1,j}+\frac{1}{z_j}\tilde\phi_{2,j}+\frac{\mu_j Q}{z_j(\tilde\psi_{1,j}-\frac{z_j\tilde\psi_{2,j}}{R^{(-1)}})}.
  \end{align}
\end{subequations}
Its Lax representation is
\begin{subequations}
  \begin{align}
    &E\psi=U\psi,\\
    &\psi_t=(V_m+\sum_{j=1}^NX_j)\psi,
  \end{align}
\end{subequations}
where $V_m=(\Gamma z^{2m})_++\left(\begin{array}{cc} 0 & 0 \\ 0 & A_{2m} \end{array}\right),$
$$X_j=\frac{1}{z^2-z_j^2}\left[\left(\begin{array}{cc} z_j^2\tilde\psi_{1,j}\tilde\phi_{1,j}^{(-1)} & zz_j\tilde\psi_{1,j}\tilde\phi_{2,j}^{(-1)} \\ zz_j\tilde\psi_{2,j}\tilde\phi_{1,j}^{(-1)} & z^2\tilde\psi_{2,j}\tilde\phi_{2,j}^{(-1)} \end{array} \right)+(\tilde\psi_{1,j}\tilde\phi_{1,j}^{(-1)}+\tilde\psi_{2,j}\phi_{2,j}^{(-1)})\left(\begin{array} {cc} \frac{z^2-3z_j^2}{4} & 0 \\ 0 & \frac{z_j^2-3z^2}{4} \end{array}\right)\right].$$

When we take $\mu_j=0$, the generalized system \eqref{Al-system-1} will reduce to the Ablowitz-Ladik hierarchy with self-consistent sources
(corresponding to the generating matrix $\Gamma$ in the power series of $1/z$).

\subsection{$\Gamma$ expanded in power series of $z$}
Similarly, Assume $\Gamma$ in \eqref{AL-gamma} as
\begin{equation}
  \bar{\Gamma}=\left(\begin{array}{cc} \bar{A} & \bar{B} \\ \bar{C} & -\bar{A} \end{array}\right)=\sum_{i=0}^{\infty}\left( \begin{array}{cc} \bar{A}_{2i}z^{2i} & \bar{B}_{2i+1}z^{2i+1} \\ \bar{C}_{2i+1}z^{2i+1} & -\bar{A}_{2i}z^{2i} \end{array}\right),
\end{equation}
then we have the following recursion relations:
\begin{subequations}
  \begin{align}
    &\bar{A}_0^{(1)}-\bar{A}_0=0,~~\bar{B}_1^{(1)}=-Q(\bar{A}_0^{(1)}+\bar{A}_0),~~\bar{C}_1=-R(\bar{A}_0^{(1)}+\bar{A}_0),\\
    &\bar{A}_{2i}^{(1)}-\bar{A}_{2i}=Q\bar{C}_{2i-1}^{(1)}-R\bar{B}_{2i-1}=Q\bar{C}_{2i+1}-R\bar{B}_{2i+1}^{(1)},\\
    &\bar{B}_{2i+1}^{(1)}=-Q(\bar{A}_{2i}^{(1)}+\bar{A}_{2i})+\bar{B}_{2i-1},\\
    &\bar{C}_{2i+1}=-R(\bar{A}_{2i}^{(1)}+\bar{A}_{2i}+\bar{C}_{2i-1}^{(1)},~~i=1,2,\cdots.
  \end{align}
\end{subequations}
where the initial value is chosen as $\bar{A}_0=\frac{1}{2}$.

The Ablowitz-Ladik hierarchy can be written in the Hamiltonian form\cite{ALzeng}
\begin{equation}
  \left(\begin{array}{c} Q \\ R \end{array}\right)_{t_m}=J^{(AL)}\frac{\delta \bar{H}_m}{\delta u}=\bar{K}^{(AL)}\frac{\delta \bar{H}_{m-1}}{\delta u}
\end{equation}
where $\bar{H}_m=-\frac{\bar{A}_{2m}}{m}$, $J^{(AL)}$ is given by (\ref{al-ham-1}), 
%%Lin and the Hamiltonian operators are
\begin{subequations}\label{al-ham-2}
  \begin{align}
%%Lin    &J^{(AL)}=\left(\begin{array}{cc} 0 & 1-RQ \\ RQ-1 & 0 \end{array}\right),\\
    &\bar{K}^{(AL)}=\left(\begin{array}{cc} \bar{K}_{11} & \bar{K}_{12} \\ \bar{K}_{21} & \bar{K}_{22} \end{array}\right),\\
    &\bar{K}_{11}=-QD^{-1}QE(1-RQ)-(1-RQ)Q^{(1)}D^{-1}Q,\notag\\
    &\bar{K}_{12}=E^{(-1)}(1-RQ)+QD^{-1}RE^{(-1)}(1-RQ)+(1-RQ)Q^{(-1)}D^{-1}R,\notag\\
    &\bar{K}_{21}=-E(1-RQ)+RD^{-1}Q^{(1)}E^{(2)}(1-RQ)+(1-RQ)R^{(1)}D^{-1}EQ,\notag\\
    &\bar{K}_{22}=-RD^{-1}R^{(1)}(1-RQ)-(1-RQ)R^{(1)}D^{-1}ER.\notag
  \end{align}
\end{subequations}
So we have another type of the GKD of Ablowitz-Ladik hierarchy
\begin{subequations}
  \begin{align}
    &\left(\begin{array}{c} Q \\ R \end{array}\right)_{t_m}=J^{(AL)}\left(\begin{array}{c} \delta H_m/\delta Q-\sum_{j=1}^N\bar\psi_{2,j}\bar\phi_{1,j} \\ \delta H_m/\delta R-\sum_{i=1}^N\bar\psi_{1,j}\bar\phi_{2,j} \end{array}\right),\\
    &(\bar{K}^{(AL)}-\frac{1}{z_j^2}J^{(AL)})\left(\begin{array}{c} \bar\psi_{2,j}\bar\phi_{1,j} \\ \bar\psi_{1,j}\bar\phi_{2,j} \end{array}\right)=0.
  \end{align}
\end{subequations}
We assume that
\begin{subequations}\label{al-constraint-2}
  \begin{align}
    &E\bar\psi_{1,j}=z_j\bar\psi_{1,j}+Q\bar\psi_{2,j}+\bar{f}_{1,j},\\
    &E\bar\psi_{2,j}=R\bar\psi_{1,j}+\frac{1}{z_j}\bar\psi_{2,j}+\bar{f}_{2,j},\\
    &E^{(-1)}\bar\phi_{1,j}=z_j\bar\phi_{1,j}+R\bar\phi_{2,j}+\bar{g}_{1,j},\\
    &E^{(-1)}\bar\phi_{2,j}=Q\bar\phi_{1,j}+\frac{1}{z_j}\bar\phi_{2,j}+\bar{g}_{2,j}.
  \end{align}
\end{subequations}
Similarly with the case of $\Gamma$ expanded in power series of $1/z$, with some simplification we assume that
\begin{subequations}
  \begin{align}
    &\bar{f}_{1,j}=zQ\bar{f}_j^2,\\
    &\bar{g}_{1,j}=zR\bar{g}_j^2,
  \end{align}
\end{subequations}
then we can find that
\begin{subequations}
  \begin{align}
    &\bar{f}_{2,j}=\frac{\mu_j}{\bar\phi_{2,j}+\frac{\bar\phi_{1,j}}{z_jR^{(1)}}},\\
    &\bar{g}_{2,j}=\frac{\mu_j}{\bar\psi_{2,j}+\frac{\bar\psi_{1,j}}{z_jQ^{(1)}}},\\
  \end{align}
\end{subequations}
where $\mu_j$ are some arbitrary constants.
So another type of the GKD of the Ablowitz-Ladik hierarchy is
\begin{subequations}\label{Al-system-2}
  \begin{align}
    &\left(\begin{array}{c} Q \\ R \end{array}\right)_{t_m}=J^{(AL)}\left(\begin{array}{c} \delta H_m/\delta Q-\sum_{j=1}^N\bar\psi_{2,j}\bar\phi_{1,j} \\ \delta H_m/\delta R-\sum_{i=1}^N\bar\psi_{1,j}\bar\phi_{2,j} \end{array}\right),\\
    &E\bar\psi_{1,j}=z_j\bar\psi_{1,j}+Q\bar\psi_{2,j}+\frac{\mu_jz_jQ}{\bar\phi_{2,j}+\frac{\bar\phi_{1,j}}{z_jR^{(1)}}},\\
    &E\bar\psi_{2,j}=R\bar\psi_{1,j}+\frac{1}{z_j}\bar\psi_{2,j}+\frac{\mu_j}{\bar\phi_{2,j}+\frac{\bar\phi_{1,j}}{z_jR^{(1)}}},\\
    &E^{(-1)}\bar\phi_{1,j}=z_j\bar\phi_{1,j}+R\bar\phi_{2,j}+\frac{\mu_jz_jR}{\bar\psi_{2,j}+\frac{\bar\psi_{1,j}}{z_jQ^{(1)}}},\\
    &E^{(-1)}\bar\phi_{2,j}=Q\bar\phi_{1,j}+\frac{1}{z_j}\bar\phi_{2,j}+\frac{\mu_j}{\bar\psi_{2,j}+\frac{\bar\psi_{1,j}}{z_jQ^{(1)}}}.
  \end{align}
\end{subequations}
Its Lax representation is
\begin{subequations}
  \begin{align}
    &E\psi=U\psi,\\
    &\psi_t=(\bar{V}_m+\sum_{j=1}^NX_j)\psi,
  \end{align}
\end{subequations}
where $\bar{V}_m=(\bar{\Gamma} z^{-2m})_-+\left(\begin{array}{cc} -\bar{A}_{2m} & 0 \\ 0 & 0 \end{array}\right),$
$$X_j=\frac{1}{z^2-z_j^2}\left[\left(\begin{array}{cc} z_j^2\bar\psi_{1,j}\bar\phi_{1,j}^{(-1)} & zz_j\bar\psi_{1,j}\bar\phi_{2,j}^{(-1)} \\ zz_j\bar\psi_{2,j}\bar\phi_{1,j}^{(-1)} & z^2\bar\psi_{2,j}\bar\phi_{2,j}^{(-1)} \end{array} \right)+(\bar\psi_{1,j}\bar\phi_{1,j}^{(-1)}+\bar\psi_{2,j}\bar\phi_{2,j}^{(-1)})\left(\begin{array} {cc} \frac{z^2-3z_j^2}{4} & 0 \\ 0 & \frac{z_j^2-3z^2}{4} \end{array}\right)\right].$$

When we choose $\mu_j=0$, the generalized system \eqref{Al-system-2} will reduce to the Ablowitz-Ladik hierarchy with self-consistent sources (corresponding to the generating matrix $\Gamma$ in the power series of $z$).
So we get two types of generalized Kupershmidt deformation of the Ablowitz-Ladik hierarchy according to the two types of Hamiltonian operators \eqref{al-ham-1} and \eqref{al-ham-2}.

\section{Conclusion}
In this paper, we construct some new integrable discrete systems
 and their Lax representations
 by making use of the generalized Kupershmidt deformation
of bi-Hamiltonian systems.
 The
 generalized Kupershmidt deformation for the Toda hierarchy,
the Kac-van Moerbeke hierarchy and the Ablowitz-Ladik hierarchy are studied.
We find that these new discrete systems
have some relations to the
equation with self-consistent sources
\cite{Mel'nikov-1983,Mel'nikov-1987,KdVWS,Liu}.
It is shown that the method can be used
to construct new discrete integrable
systems in $(1+1)$-dimensional case.
 The
Kupershmidt deformation for higher dimensional case still needs to be studied.

\section*{Acknowledgments}
The authors are grateful for referee's valuable comments to improve
our manuscript.  %Linarticle.
This work was supported by the National Science Foundation of China
(Grant no. 10901090, 11171175, 11201477) and Chinese Universities
Scientific Fund (2011JS041). %Lin The first author
Y. Huang is also supported by Special Funds for Co-construction Project of Beijing.

% ------------------------------------------------------------------------
\end{document}